%% file: main.tex
\documentclass{jot}
\PassOptionsToPackage{table}{xcolor}
\PassOptionsToPackage{xcdraw}{xcolor}
\usepackage{afterpage}
\usepackage{multirow}
\usepackage{subcaption}
\usepackage{setspace}

\jotdetails{
    volume=25,      % volume
    number=3,       % number or issue
    articleno=a18,   % article number, eg a1 for research articles, e for editorials
    year=2026,      % year
    license=ccbyncnd    % choose from ccby, ccbynd, ccbyncnd
}

\articletype{regular}
\title{A Model-Based Framework for Developing DTs in Industry 4.0}

\author[$\ast$]{Lina Bilal}
\author[$\ast$]{Benoit Combemale}
\author[$\ast$]{Jean-Marc Jézéquel}
\author[$\S$]{Quentin Perez}

\affil[$\ast$]{Inria, IRISA, CNRS, University of Rennes, France}
\affil[$\S$]{INSA Rennes, Inria, IRISA, CNRS, University of Rennes, France}

\runningtitle{A Model-Based Framework for Developing DTs in Industry 4.0} % For use in the internal pages 

\runningauthor{Bilal \textit{et al.}}

\input{src/01.abstract}

\usepackage[most]{tcolorbox}

\begin{document}
\newtcolorbox{infobox}{
  colback=gray!10,
  colframe=black,
  arc=3pt,
  boxrule=2pt,
  left=6pt,
  right=6pt,
  top=6pt,
  bottom=6pt
}

\maketitle
\urlstyle{rm}
\input{src/02.introduction}

\input{src/03.motivatingExample}
\input{src/04.approach}
\input{src/05.implementation}

\input{src/06.evaluation}

\input{src/07.relatedwork}
\input{src/08.conclusion}

\input{src/09.acknowledgments}

%\begin{figure}[h!]
   % \centering
  %  \fbox{\includegraphics[width=0.95\columnwidth]{figures/jot.png}}
 %   \caption{The title page of a JOT manuscript.}\label{fig:titlepage}
%\end{figure}

% \section{Author short bio}
% A short bio of the authors (in the same order as in the title) must be included at the end of the paper, after the references, as follows:

% where \command{authorcontact} is used to provide the \param{homepage} (optional) and the \param{email}.

\bibliography{bibliography}
\section*{About the authors}
\shortbio{Lina Bilal}{is a PhD student at the University of Rennes. She earned her Engineering degree in Computer Science at ENSA Oujda (Morocco), followed by a Master’s degree in Software Engineering at Aix-Marseille University. Her research interests focus on the topics of Model-Based DevOps, Model-Driven Engineering and Digital Twins. \authorcontact{lina.bilal@irisa.fr}}
\shortbio{Benoit Combemale}{is a Research Director at Inria and a Full
Professor of Software Engineering at the University of Rennes.
His research interests in Software Engineering include Software
Language Engineering, Model-Driven Engineering, and Software Validation \& Verification. \authorcontact[https://people.irisa.fr/Benoit.Combemale/]{benoit.combemale@irisa.fr}}
\shortbio{Jean-Marc Jézéquel}{is a Professor of Software Engineering at the University of Rennes and a member of the DiverSE team at IRISA/Inria, as well as a fellow of the Institut Universitaire de France (IUF). Since 2024 he is President of Informatics Europe. From 2012 to 2020 he was Director of IRISA, one of the largest public research lab in Informatics in France. In 2016 he received the Silver Medal from CNRS and in 2020 the IEEE/ACM MODELS career award. He was an invited professor at McGill University in 2022.\authorcontact[https://people.irisa.fr/Jean-Marc.Jezequel/]{jean-marc.jezequel@irisa.fr}}
\shortbio{Quentin Perez}{is an Associate Professor at INSA Rennes and a member of the DiverSE team at IRISA/Inria. His research interests in Software Engineering concern Empirical Software Engineering, Green Software Engineering and Digital Twins. \authorcontact[https://qperez.github.io/]{quentin.perez@irisa.fr}}
%\onecolumngrid
\end{document}

%% file: src/01.abstract.tex
\begin{abstract}

With the rise of Industry 4.0 driven by the integration of Cyber-Physical Systems (CPS) and the Internet of Things (IoT), the use of Digital Twins (DTs) has significantly increased over the past decade, as they provide detailed insights and support well-informed decision-making. However, the lack of standardized methodologies, in addition to the time and resources involved for building them remains an important challenge.

Building on the idea that engineering models of the physical twin (PT) are often available, we propose a tool-supported framework that automates the derivation of DTs by leveraging existing structural and behavioral models of the PT and extending them with additional models to build a comprehensive DT. To demonstrate the feasibility of our approach, we applied it to four different use cases, in which we automatically derived DT instances from (1) models of their PT, (2) configuration of our generic framework and (3) minimal ad hoc additional development for connecting the DT to the PT. These experiments illustrate the applicability of our framework for building DTs in contexts that satisfy our assumptions and requirements. By simply configuring the framework, we are able to derive a DT aligned with its operational purpose.

\end{abstract}

%% file: src/02.introduction.tex
\section{Introduction}

The emergence of Industry 4.0 has revolutionized industrial and technological fields through the integration of cyber physical systems (CPS) and the internet of things (IoT). In this context, the concept of Digital Twins (DTs) has become a key technology for linking physical and digital entities. While originally conceptualized with NASA’s Apollo 13 mission \cite{glaessgen2012nasa}, Grieves formally defined it as ‘Mirrored Spaces Model’ \cite{grieves2005paradigm} and later as ‘Information Mirroring Model’ \cite{grieves2005plmlean}. Over time, their application has expanded across various domains, including manufacturing, healthcare, agriculture, thus enhancing understanding of their physical counterpart's behavior, and the capacity of making well informed decisions.

As DTs become more pervasive, there is a growing need for their rapid development \cite{rasheed2020dt,tao2019sota}, for they help reducing time and costs in manufacturing \cite{tao2019sota}. 
However, despite its advantages, the development process of a DT remains complex due to a number of challenges \cite{bordeleau}.

The main challenge for implementing a DT comes from the complexity involved in the modeling and development stages. This is due to the involvement of different stakeholders and the need to address multiple concerns throughout both the design and operational phases. The creation of a DT is both time and resource consuming, encompassing domain knowledge and the selection of appropriate models regarding granularity and scope.

Another key challenge is the lack of standardized methodologies for the design and implementation of DTs \cite{oakes2024edt,judith2024mbedt,bibow2020injection,munoz2021,kirchhof2020,chakraborty2021surrogate}. The lack of consistent guidelines or accepted protocols to guide their development makes it difficult for industries to adopt and deploy DT solutions. 

%-Q.P: automated or semi-automated? 
Building on the observation that engineering models of the physical twin (PT) are often readily available in Industry 4.0 settings, this paper proposes an automated approach to DT development. By leveraging existing PT models, the proposed approach establishes a systematic methodology that mitigates the complexity and resource demands typically associated with DT implementation.

The main contribution of this paper is a tool-supported approach for the automated derivation of DTs in Industry 4.0 environments. The approach initially leverages on the design models of the physical twin to derive core DT aspects. Subsequently, additional models are identified and integrated to enable the construction of a comprehensive and coherent DT.

The remainder of this paper is structured as follows: Section 2 provides the necessary background. Section 3 motivates our work with a motivating example. Section 4 presents an overview of our proposed tool-supported framework. Section 5 details the implementation choices of the framework. Section 6 discusses the evaluation and results of the proposed approach. Section 7 reviews related work. Section 8 concludes the paper.

%% file: src/03.motivatingExample.tex
\section{Motivating Example} \label{sec:ill}
In this paper, we set ourselves in a situation where the PT and its DT are developed independently and follow different lifecycles. The PT is considered to be already functional prior to the integration of its DT.
\subsection{Physical Twin}
We illustrate our approach with a Vacuum Gripper Robot (VGR) (Figure \ref{vgrimg}) of the Fischertechnik\footnote{\url{https://www.fischertechnik.de/en}} factory, a 3-axis robot serving to position workpieces in a three-dimensional space. It operates along horizontal, vertical and rotational movements, with each axis actuated by a motor equipped with an encoder, which allows tracking the robot's movement along that axis, and a limit switch that detects end positions. The robot includes a vacuum based gripper composed of a compressor and a valve to pick and place objects. 
These components are orchestrated by a Programmable Logic Controller (PLC), which is responsible for reading sensor inputs and issuing actuator commands through digital input/output operations. The VGR provides a predefined set of commands that can be issued to perform specific operations, such as picking, placing, moving to a target position or returning to a reference position. At the gateway level, system state and data are published to an MQTT\footnote{\url{https://mqtt.org/}} broker, while control commands are received via WebSocket, allowing external software to interact with the VGR.
The VGR also features a telemetry layer, to allow reporting detailed runtime information beyond what is strictly necessary for its primary operation. 
This integration between mechanical components, sensing and control logic makes the VGR a representative system for exploring smart manufacturing in the context of industry 4.0. 
\afterpage{
\begin{figure}[htbp]
\centerline{\includegraphics[width=1\columnwidth]{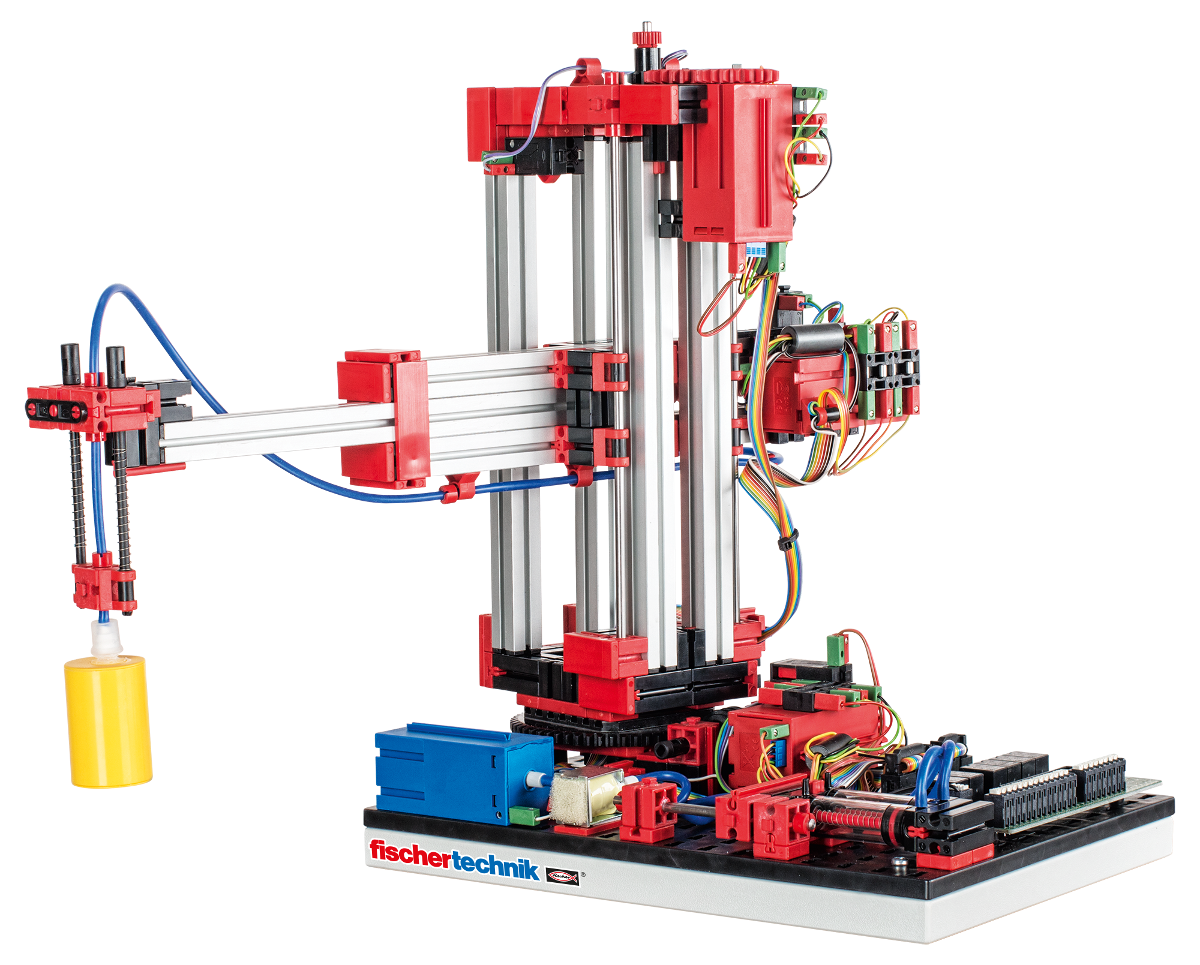}}
\caption[]{Vacuum Gripper Robot\footnote{\url{https://www.fischertechnik.de/en/products/industry-and-universities/training-models/536630-vacuum-gripper-robot-24v}} borrowed from the Fischertechnik factory.}
\label{vgrimg}
\end{figure}
}
\subsection{Models of the Physical Twin} \label{sect:mopt}
In this paper, we assume that in an industrial context, we would already have the models that provide the necessary blueprints to build the robot. Since we started with an already existing robot, we reverse engineered the VGR based on its specifications and available documentation to create detailed SysMLv2 models of the PT, capturing both structural and behavioral aspects.

\begin{figure}[htbp]
\centerline{\includegraphics[width=0.8\columnwidth]{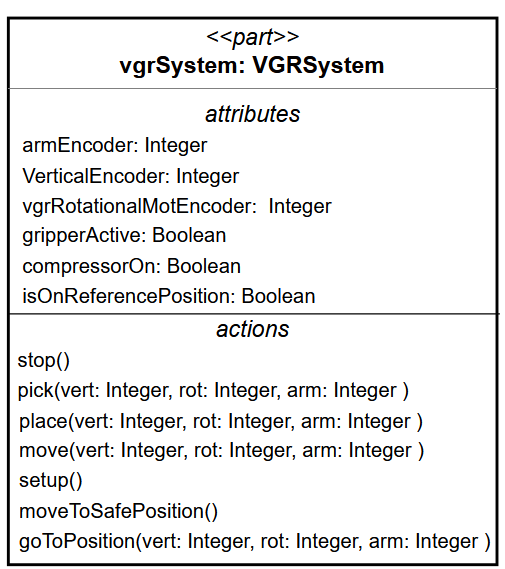}}
\caption{Excerpt of the Conceptual Block diagram of the VGRSystem.}
\label{vgrsystem}
\end{figure}

To structurally represent the VGR, we adopted a hierarchical decomposition approach, following a top-down strategy. For the purpose of this work, we focus on the conceptual level, which provides a high-level overview of the system and its exposed attributes (cf. Figure \ref{vgrsystem}). This abstract representation hides away any internal complexity, such as electrical wiring and pneumatic connections, and provides a holistic view of the VGR system.

Figure \ref{vgrsystem} illustrates an excerpt of a conceptual SysML V2 model of the VGRSystem, highlighting some of its exposed attributes. Attributes such as \emph{armEncoder} represent encoder values that indicate the current position of the robot along the corresponding axis. These attributes reflect externally visible information, abstracting away the internal details about how these values were obtained.

\begin{figure*}[htpb]
\centerline{\includegraphics[width=0.9\textwidth]{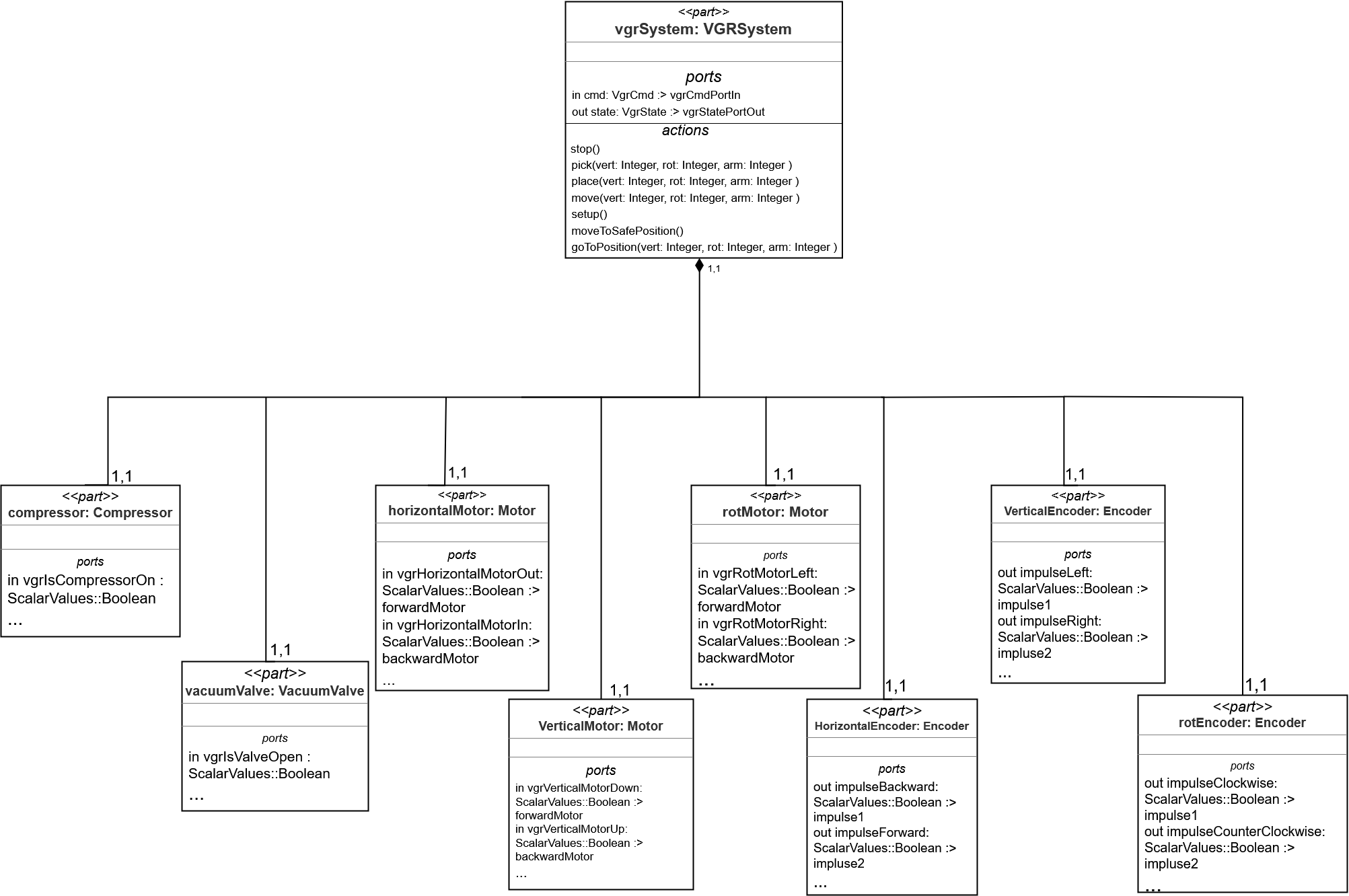}}
\caption{Excerpt of a Detailed Part Definition Diagram of the VGRSystem.}
\label{detailedVGR}
\end{figure*}

At a more detailed level, the VGR is decomposed into its different components (cf. Figure \ref{detailedVGR}). This level provides a more concrete structural representation of the system, describing how components are connected and their interactions with one another. It includes internal aspects, such as directional motor commands and encoder impulses, which explain how the high-level values observed at the conceptual level were produced.

\begin{figure}[htpb]
\centerline{\includegraphics[width=1\columnwidth]{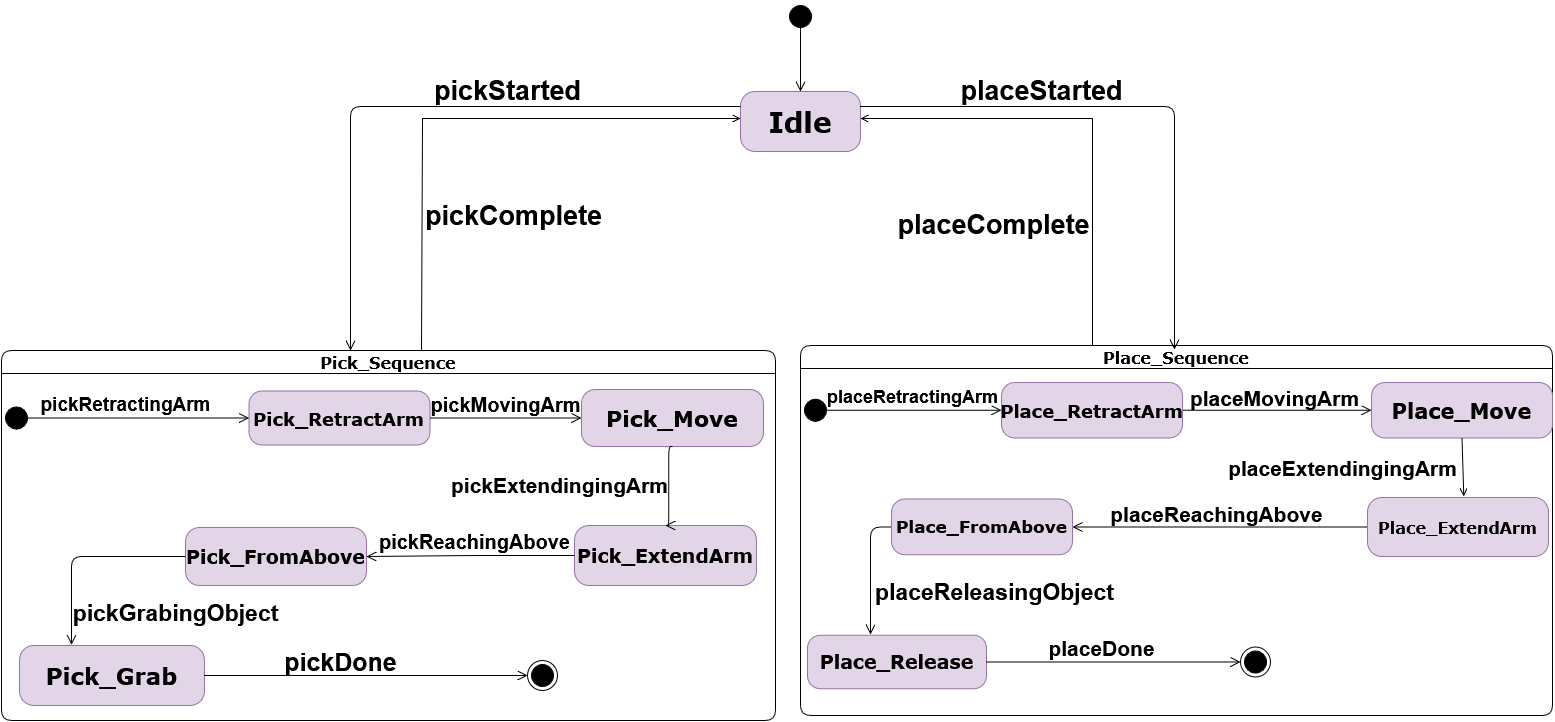}}
\caption{Mission model of the VGR.}
\label{behav}
\end{figure}

Behavioral models represent the system's progression through state transitions (e.g., from idle states through execution phases), performing actions like moving to positions, picking and dropping objects. In our case, we only focus on deterministic execution paths, to ensure a verifiable and predictable execution of the robot's missions. 
These models represent the robot's mission (cf. Figure \ref{behav}) that informs the logic and behavior that will be deployed on the PT. 
% They serve as the basis for reasoning over expected behaviors of the PT, supporting real-time comparison with sensor feedback. 

Figure \ref{behav} presents the statechart modelling the mission  of the VGR. It performs pick and place sequences, referring respectively to the actions of taking an object with the gripper at a specific pick-up location and releasing the gripper to place the object at a designated drop-off location. Each sequence has nested states, where the pick sequence involves retracting the robotic arm to a reference position, moving it to the pickup position, extending the arm and activating the gripper.
The place sequence follows a similar structure, ending with gripper deactivation and object release.

\subsection{Digital Twin}\label{ss:dt}
We aim to develop a DT of the VGR with the objective of supporting basic visualization via a dashboard, detecting anomalies in the behavior of the VGR and controlling it via a cockpit.
For that, a DT is made of several software components, as illustrated in Figure \ref{components}. At its core are models that virtually replicate the geometric and physical properties of the PT, capturing its structure, function, and environment. They encompass descriptive models that represent current or past states to facilitate understanding and analysis of the PT, predictive models that forecast future or unmeasured behaviors and prescriptive models that define intended behaviors and guide system interventions \cite{benoit2021models}. These models rely on continuous streams of data collected from sensors on the PT, establishing real time synchronization \cite{eramo2021conceptualizing}. Building on this, services leverage both models and real-time data required by the DT. 
% \begin{figure}[htpb]
% \centerline{\includegraphics[width=\columnwidth]{figures/dt-components.png}}
% \caption{Physical Twin and Digital Twin Components for the VGR.}
% \label{components}
% \end{figure}

The gateway of the DT handles the bidirectional communication between the two entities. 
Its design depends on how the PT and DT are integrated. If both twins are co-developed, the gateway can directly access the PT without additional specification. In a more decoupled setting, where the twins evolve independently, such as in our motivating example, the PT must expose an interface to allow external access to the DT.
The gateway's role is to ensure data flow through defined communication protocols, allowing the DT to receive real-time updates from the PT and for the DT to send commands to close the feedback loop, thereby maintaining synchronization between the two entities.

\subsection{Models of the Digital Twin}
 
Models of the DT refer to architectural representations that describe the structure, behavior and services of the digital twin \cite{micheal2025mdedtops}. They capture how the DT is organized, how its services interact and how it communicates with its PT.

\begin{figure}[h]
\centerline{\includegraphics[width=1\columnwidth]{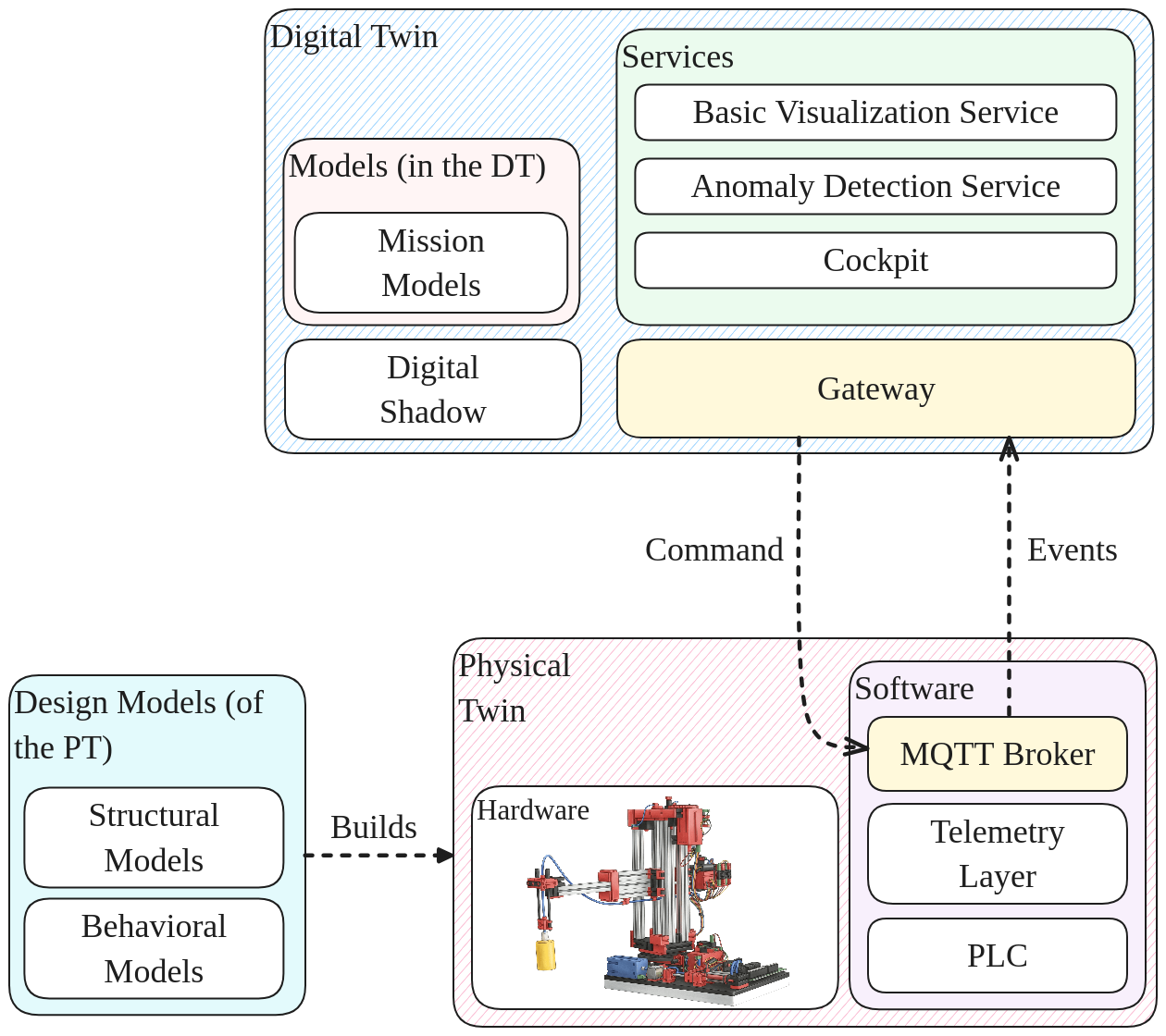}}
\caption{Physical Twin and Digital Twin Components for the VGR.}
\label{components}
\end{figure}

To develop a DT for the VGR, we identified which functionalities it should offer, including basic visualization, detecting anomalies and exposing commands to be sent to the VGR via a cockpit. Each of these functionalities corresponds to a specific service that must be explicitly represented in the DT's architecture (cf. Figure~\ref{components}).

These models served as a foundation for identifying the required components and specifying how they interact and exchange information. We define the main elements of the DT, including the gateway for managing communication, the digital shadow for storing incoming data and the services for fulfilling the DT's objectives.

We should also be able to define how the DT should communicate with the VGR. Since the VGR exposes its data over MQTT, we explicitly configure the connection to the MQTT broker, define how to handle incoming messages and filter them to get only the data relevant to the DT.
%and define how to handle and parse incoming messages.

These models of the DT serve as a basis for developing the DT of the VGR that can operate in real-time and maintain synchronization with its physical counterpart.
%define the DT's architecture, including the components it's composed of, the services it offers and the way it processes data. 

\subsection{Models in the Digital Twin}
Models in the DT refer to representations of the PT that are embedded within the DT and used at runtime to support the services it provides \cite{micheal2025mdedtops}. These models may serve descriptive, prescriptive and predictive purposes and are used to help the DT interpret or reason over the PT's behavior.
Each service relies on a specific type of model. For the VGR, the anomaly detection service requires comparing the actual behavior of the VGR with the expected behavior. To achieve this, the DT needs a model that describes precisely how the mission is supposed to be executed by the PT.  
The models in the DT are thus essential for allowing the DT to fulfill its objectives, as they provide necessary information to interpret, evaluate and act upon the VGR (cf. Figure~\ref{components}).

% \begin{figure}[htbp]
% \centerline{\includegraphics[width=\columnwidth]{figs/DTComp-br.png}}
% \caption{DT Components}
% \label{components}
% \end{figure}

\begin{infobox}
Our motivating example reveals the importance of distinguishing between 3 types of models. The first are {\bf models of the PT} that describe and represent the PT. The second are {\bf models of the DT} that define the DT's architecture, its internal components and how they interact with one another.
The third are {\bf models in the DT} that are embedded in the DT to support the DT's functionality at runtime.
\end{infobox}
\subsection{Stakeholders}\label{stk-ill}

\begin{figure*}[htbp]
\centerline{\includegraphics[width=0.8\linewidth]{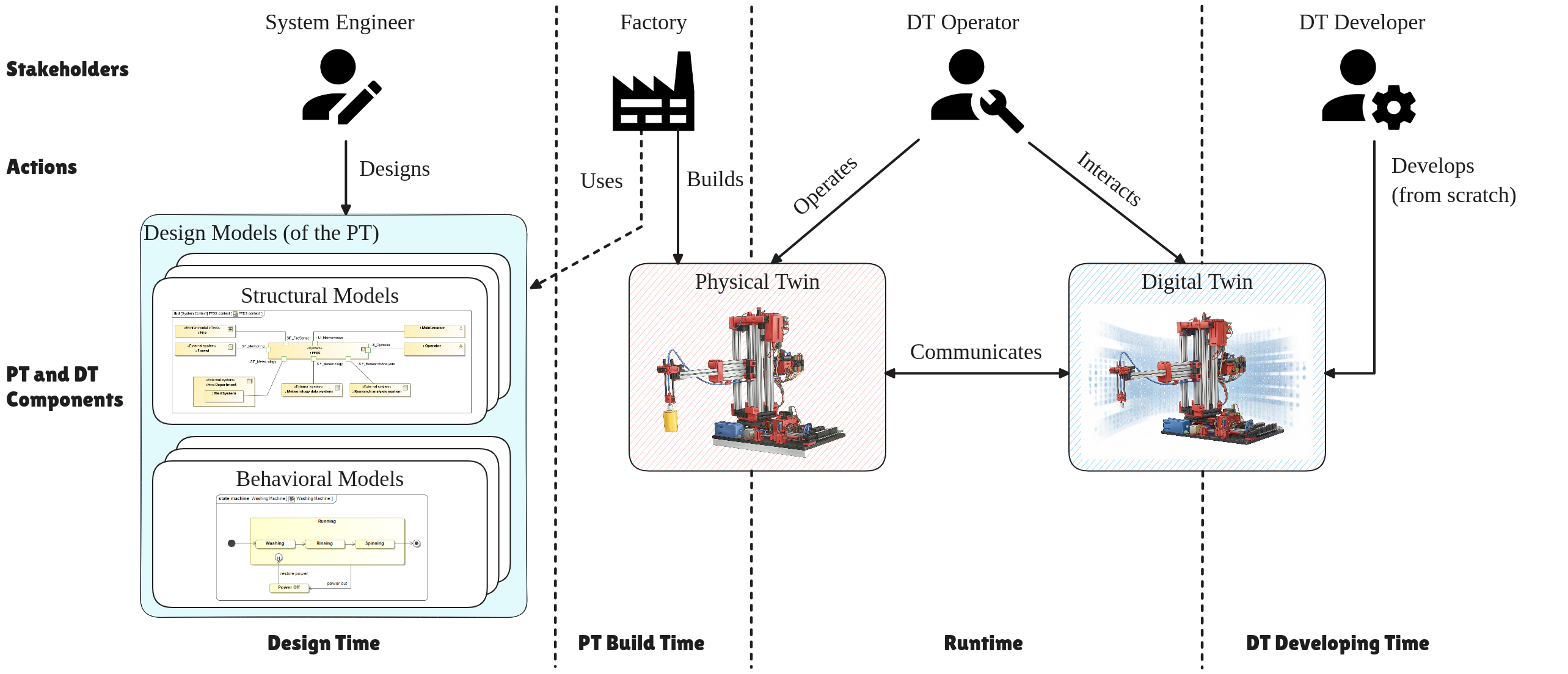}}
\caption{Roles interacting with PT and DT}
\label{oldstk}
\end{figure*}

The development of these components needs a close collaboration between different roles, as illustrated by Figure \ref{oldstk}: 
\begin{itemize}
    \item \textbf{Systems engineer} responsible for creating detailed models of the PT, capturing structural elements, behavior and interconnections between the different component at design time. They make sure to provide a clear documentation of the VGR's intended functionality. These models are the blueprints for building the PT.
    \item \textbf{DT developer} plays a dual role, combining the skills of a software engineer, responsible for the architecture and technical infrastructure of the DT and of a domain expert who brings the knowledge needed to accurately model the VGR. The DT developer oversees the entire lifecycle of the DT, including its development, deployment, maintenance and ongoing evolution and must consider multiple concerns in building the DT. First, they need to create representative models of the VGR and handle incoming real-time data. The DT must be designed to fulfill its purpose, which requires translating its intended functionalities into dedicated services, that are basic visualization, detecting anomalies and controlling the VGR via a cockpit. As the VGR changes, the DT developer is responsible that the DT evolves with the changes to its physical counterpart.
    Additionally, they need to establish a bidirectional connection through communication protocols such as MQTT or OPC-UA, enabling the DT to send control actions to the PT, and thus closing the feedback loop. 
    
    \item \textbf{DT operator} once DT is operational, the DT operator relies on the services of the DT, to continuously monitor the VGR, and assess its performance. They interpret alerts and identify anomalies and could determine the necessary corrective actions for them. 
\end{itemize}

% \begin{figure}[htbp]
% \centerline{\includegraphics[width=\columnwidth]{figs/illustr-ex-roles.png}}
% \caption{Roles interacting with PT and DT}
% \label{fig}
% \end{figure}

\subsection{Challenges}

Currently, developing such a DT from scratch is challenging due to the manual complexity involved in the different concerns the developer encounters. These concerns include system modeling, handling different data streams, implementing all the services required to fulfill the DT's objective, and closing the feedback loop by sending commands back to the PT.

In developing the DT for PT such as the VGR, we have identified a set of recurring needs.
%that highlight essential requirements for a DT framework.
We aim at targeting industrial systems that expose their data through standard communication protocols such as MQTT or OPC-UA, and rely on deterministic state-based behaviors that can be monitored at runtime.

In this context, DTs often share a similar architecture, with common components such as gateways, digital shadows and services. Rather than redefining this architecture from scratch for every new DT, we propose to establish a generic and reusable architecture, that can then be adapted as needed.

We also found that some of the runtime models required by DT services, such as the mission model used for anomaly detection, had already been defined during the design phase of the VGR. This suggests that we can reuse models of the PT as models in the DT, to avoid unnecessary redundancy and to save modeling efforts. 

Nevertheless, even with generic architecture and reused models, each DT still needs to be tailored to its purpose and specific characteristics of the PT it represents. The framework must therefore support configuration, allowing the DT to be adapted according to its intended purpose and to the specific requirements of its physical counterpart.

These considerations lead us to address the following research questions (RQ).

\noindent\textbf{RQ: How can we provide a framework to support model-based development of DTs in Industry 4.0?}
\newline
\noindent This RQ is explored through four sub-questions.
\newline\newline
\noindent\textbf{$\circ$ RQ1}: What elements of the models for the PT can be used for deriving parts of the DT?

While they provide a strong foundation by accurately representing the PT and its behavior, they are probably not sufficient on their own to fully generate a functional DT. Additional models of the DT are required to capture the full complexity of the DT.
    
\noindent\textbf{$\circ$ RQ2}: What additional models are necessary for developing a DT?

\noindent\textbf{$\circ$ RQ3}: How generic is the approach within the context of Industry 4.0?

\noindent\textbf{$\circ$ RQ4}: To what extent does the framework support the co-evolution of the PT and DT over system changes? 

%% file: src/04.approach.tex
\section{Approach Overview} \label{sec:approach}
% We propose a MDE approach targeting such a configurable framework to automate the construction of a DT. This automation could significantly reduce complexity, as configuring the framework would suffice to instantiate a fully functional DT that fulfils its intended purpose.
% The framework targets industrial systems that expose their data through standard communication protocols such as MQTT and OPC-UA and rely on deterministic state-based behaviors that can be modeled and interpreted at runtime.

% Building on this, the framework is applicable to any industrial setting that satisfies a similar set of services (cf. Sect.~\ref{sec:ill})

We propose an MDE approach targeting a configurable framework to automate the construction of a DT, that is applicable to any manufacturing setting that satisfies a similar set of services (cf. Sect. \ref{ss:dt}). This automation could significantly reduce complexity, as configuring the framework would suffice to instantiate a fully functional DT that fulfills its intended purpose.

% -- TODO: Requirements: 
% Our approach is applicable to any system following a small set of requirements. The PT must expose data coming from its different sensors that the industrial IoT engineer must expose at a specific frequency, to be used for the DT. 

\subsection{Configurable framework}
This approach relies on a fully configurable and reusable framework for developing DTs. Rather than hard coded logic, the framework is driven by the models of the PT, and the additional models of the DT for configuring the DT (Figure \ref{stk-fw}). %  (cf. Sect.~\ref{sec:ill})

% our generic and configurable DT framework, which abstratcs dt's core components and  structures them into a modular architecture that can be easily adapted to different physical systems.

At the core of the framework lies the \textbf{DTEngine}, which serves as the central coordinator. It is guided by the configuration that defined system-specific parameters. Beyond initializing the different components, the DTEngine manages data flow between them, delegating incoming data to the appropriate components and activates different services selected for the DT's purpose. 
%Service instantiation is driven by the feature model, to make sure only the selected services are activated.

A communication layer is built around the \textbf{DTGateway}, which serves as the bridge between the PT and the DT. It handles incoming data from the PT and forwards it to the DTEngine.
The gateway abstracts the details of various communication protocols, allowing the framework to support new protocols by simply extending the interface, without modifying the already existing code base. This is achieved by specifying the connection, disconnection, as well as configuring data collection, with the possibility to define additional behaviors if required by the communication protocol.

To interpret system behavior, the framework includes a \textbf{ModelInterpreter} component, that handles the parsing of structural models and the interpretation of behavioral models. It exposes model information needed by other DT components. Additionally, the interpreter analyzes behavioral elements such as states, transitions, guards and timing constraints. At runtime, it evaluates whether the DT is aligned with the PT, to validate the system's state. 

The architecture of the framework supports extensibility by allowing additional modeling languages to be integrated through new implementations that comply with the ModelInterpreter interface. In our work, we currently support SysML V1 or V2.
These implementations redefine core methods to expose structural and behavioral elements.
Depending on the modeling language, additional methods may also be necessary to handle the parsing and evaluation of guards and constraints.

\textbf{DigitalShadow} plays a central role in capturing and maintaining a time-stamped, continuously updated data of the PT's state. Real time data received through the PT is forwarded by the DTEngine. The stream of feedback is persistently stored in a time series databases, allowing both access to the real time system values, as well as access to the historical data. 
%The digital shadow thus serves as the historical memory of the DT.

On top of this infrastructure, our framework supports a set of generic services that are not hardcoded or initialized all at once. Instead, they are selected during configuration based on the intended purpose of the DT. 
%This selection is captured in the feature model, and the DT Engine initializes only the services corresponding to the chosen features.

The services currently offered by our framework are carefully selected to demonstrate different ways the system can interact with models and real-time data at runtime. They represent three complementary categories of functionality including reading and visualising collected data, interpreting and reasoning over models at runtime, and sending commands back to the PT.

The selection of these services covers the essential capabilities needed to an operational DT that are as follows: 

\begin{description}[leftmargin=1cm]

% Basic Visualization: The ability to graphically or parametrically (that is, through parameters and values) visualize data through simple charts, graphs, simple dashboards, tables, hierarchical and basic 3D views of the as sets.
%\ref{dtc-periodic-table-services}

\item[Basic Visualization Service]: as described in the Digital Twin Consortium \footnote{\url{https://www.digitaltwinconsortium.org/initiatives/capabilities-periodic-table/}}, this service provides graphical visualization of PT data through charts and graphs. In our framework, the corresponding user interface (UI) is automatically generated and it includes the set of data selected at configuration time, which are then visualized at runtime.

\item[Anomaly detection service]: plays a central role in maintaining fidelity of the DT, by continuously monitoring the system's real time behavior and comparing actual behavior of the PT against expected behaviors specified in the models in the DT. This service demonstrates the ability to reason dynamically over models at runtime. Anomalies are flagged upon detecting discrepancies such as missing, delayed or incorrect transitions.

%interacts with the appropriate service to solve it. 

% \item[Reconfiguration service]: 

% detects patterns in the system's behavior that require configuration of the PT to adapt the system to evolving conditions. Additionally, when the anomaly detection services identifies a deviation from the expected behavior, the reconfiguration service sends an appropriate corrective action to the PT.

\item[Cockpit service]: provides a UI to issue commands from the DT to its physical counterpart. These commands, their parameters and associated constraints are derived from the models of the PT and exposed through the cockpit UI and are sent to the PT through the selected communication protocol.

\end{description}
\subsection{Framework Stakeholders}
\begin{figure}[htbp]
\begin{center}
\includegraphics[width=0.9\columnwidth]{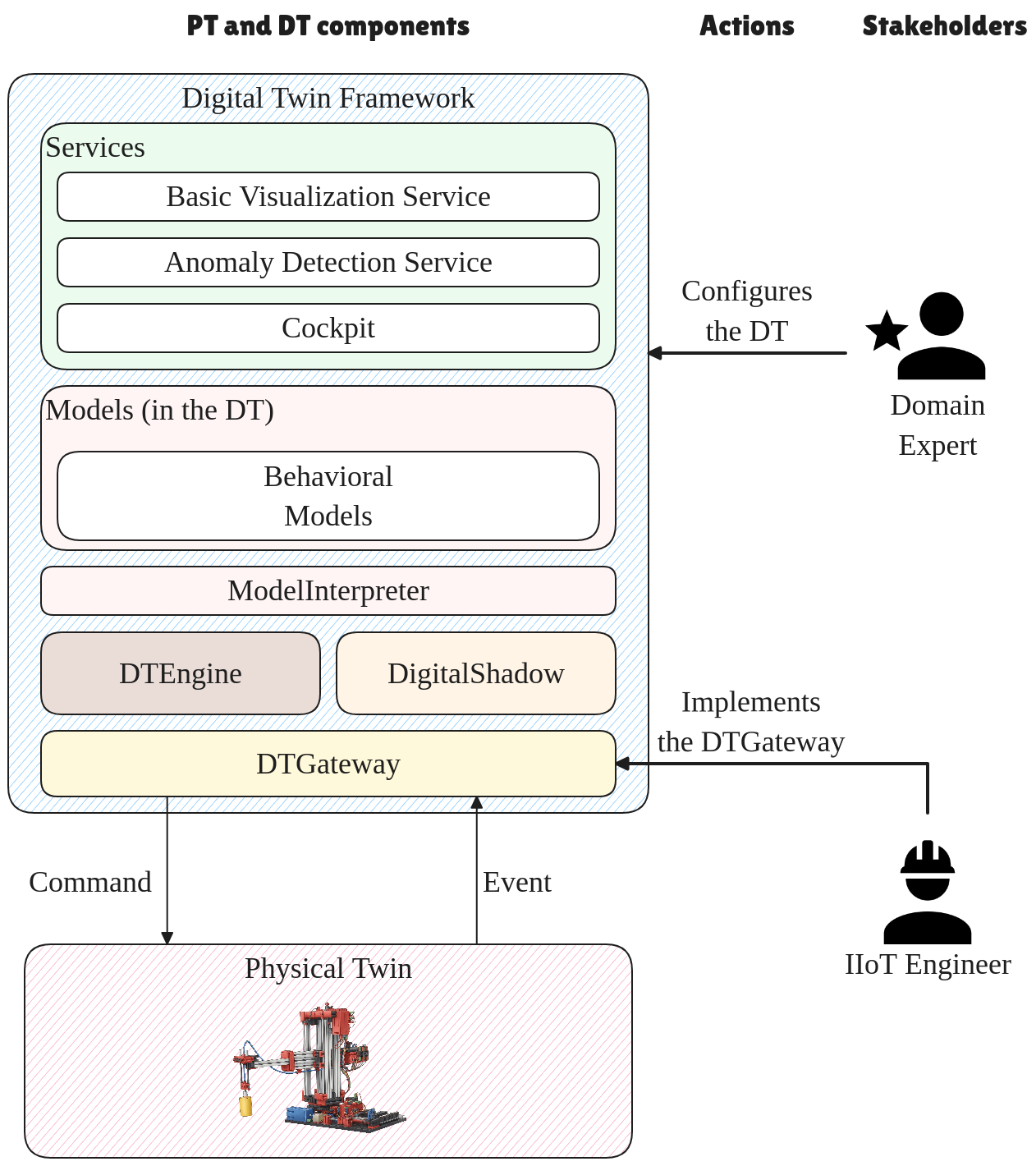}
\caption{Architecture and New Stakeholders of the Digital Twin Framework.}
\label{stk-fw}
\end{center}
\end{figure}

Our framework captures the software engineering expertise required for DT development, introducing two new roles (Figure \ref{stk-fw}) that replace the traditional stakeholders (cf. Sect \ref{stk-ill}). 

\begin{itemize}
    \item \textbf{Domain expert}: a PT expert who understands the system and the models that describe it. They define why a DT is needed by specifying its objectives and realize them by configuring the DT accordingly. They also make sure that the resulting DT instance provides an accurate representation of the PT. This role combines the responsibilities of both the Systems Engineer and the DT Operator in a traditional DT development methodology (cf. Section \ref{stk-ill}).
    \item \textbf{Industrial IoT (IIoT) engineer }: responsible for setting up the connectivity infrastructure and implementing the gateway component that links the PT and the DT. This includes capturing sensor measurements and events from the PT, and handling  necessary abstractions of the gateway such as data aggregation or complex event processing, as required by the PT to combine and publish relevant events to the DT. The IIoT engineer plays part of the DT Developer's role, focusing only on the gateway implementation.

\end{itemize}
% \begin{figure}[htbp]
% \centerline{\includegraphics[width=\columnwidth]{figs/Stakeholders.png}}
% \caption{New Stakeholders of the DT Framework.}
% \label{stk-fw}
% \end{figure}

Through this separation of concerns, domain experts can configure and operate their DT instances without software development expertise, while IIoT engineers would focus on connectivity rather than application logic.

\subsection{Framework Configuration}
% JMJ=> déjà dit ci-dessus: Our approach leverages the models of the PT to automate the development process of the DT. These models provide a structural and behavioral specification of the system under study. Our goal is to identify what aspects of the DT can be directly derived from these models, and what models of the DT need to be identified to construct a fully functional and configurable DT.

From the models of the PT, and specifically from the structural models, written in SysML v2, we extract part definitions and their associated attributes to automatically generate a database schema. This schema serves as the blueprint for data storage in the DT, defining the set of attributes to be written to the database.
The models of the PT guided the engineering of the PT, such that the data it produces matches the attributes defined in the model. This alignment allows the automatic mapping between model attributes and incoming PT data, allowing the digital shadow to access and track the real time value of each attribute directly from the incoming feedback.

We also configure the Basic Visualization service, since the data we observe are specified in these models. The same principle applies to the cockpit service, where the commands are derived from the models and are sent back to the PT through the chosen communication protocol.

From the behavioral models, we can configure the anomaly detection service aiming at verifying whether the PT and the DT are aligned. By continuously comparing the execution state of the DT against its physical counterpart, the system could detect anomalies.

Additionally, the level of abstraction captured in the models of the PT directly determines the granularity at which the DT can operate. Whether these models describe the PT at a conceptual level or provide detailed physical specifications, the corresponding DT can be instantiated accordingly. Each level of abstraction allows for a purposeful DT instantiation, adapted to the precision of the models available.

\begin{figure}[htbp]
\centerline{\includegraphics[width=1\columnwidth]{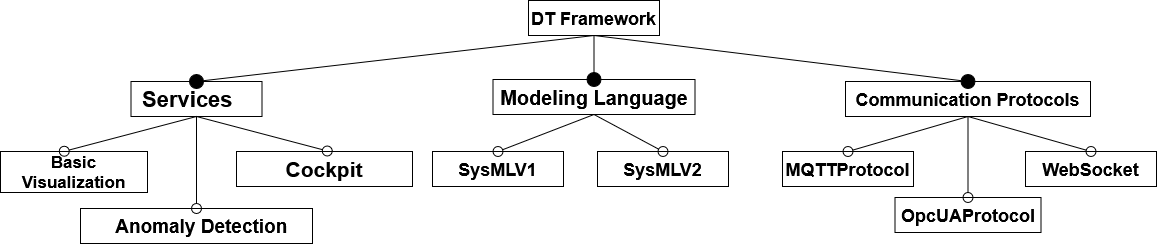}}
\caption{Feature Model illustrating the Configurable Aspects of the DT Framework.}
\label{feature}
\end{figure}

To configure our DT framework, the domain expert must specify the models of the PT as input to the framework. We propose a feature model (Figure \ref{feature}) to support the selection of the different configurable aspects of the system, including the services to activate, the communication protocol and the modeling language used to define the models of the PT, to guarantee that the DT is tailored to meet its intended purpose. 
The domain expert can then configure the framework using this feature model. Depending on the selected services, the configuration also includes selecting a subset of relevant model elements for each service. These include the data to be visualized via a dashboard and the commands to be exposed for the cockpit service.

% The domain expert can then configure the framework using this feature model, to obtain a so-called resolved model that captures parameters critical to the runtime behavior of the DT. These include a subset of data to be selected, for the different services. 

%since we do not want to receive all the data from the DT,

%It also includes data mapping from the models of the PT to the actual data coming back from the PT, to help evaluate different conditions when interpreting the models of the PT. 

%\subsection{Scope?}
%Our scope targets industrial systems that expose their data through standard communication protocols such as MQTT or OPC-UA, and rely on deterministic state-based behaviors that can be interpreted at runtime. The framework is built with these characteristics, with support for protocol-specific connectivity and runtime interpreters for behavioral models. It supports DT construction in systems that match these characteristics, while applying the same approach to other domains would require different design considerations.

%The framework targets industrial systems that expose their data through standard communication protocols such as MQTT and OPC-UA and rely on deterministic state-based behaviors that can be modeled and interpreted at runtime.

%We can also specify which communication protocol (for e.g. MQTT, OPC-UA) that handles the connectivity between the DT and its physical counterpart. It also supports the specification of the model interpreter to be used, choosing between SysML v1 and SysML v2, based on the format of the input models. 

%TODO expliquer workflow
%--> workflow figure

%% file: src/05.implementation.tex
%\section{Model-Based Framework for DT Development}
\section{Implementation choices} % JMJ could be a subsection of the previous section (?)

This section details the concrete implementation of the DT framework previously introduced (cf. Sect. \ref{sec:approach}), highlighting the technological choices and specific tools used to realize the conceptual components of the DT, from the configuration phase to the execution of the DT instance.

\begin{figure}[t]
  \centering
\includegraphics[width=\columnwidth ]{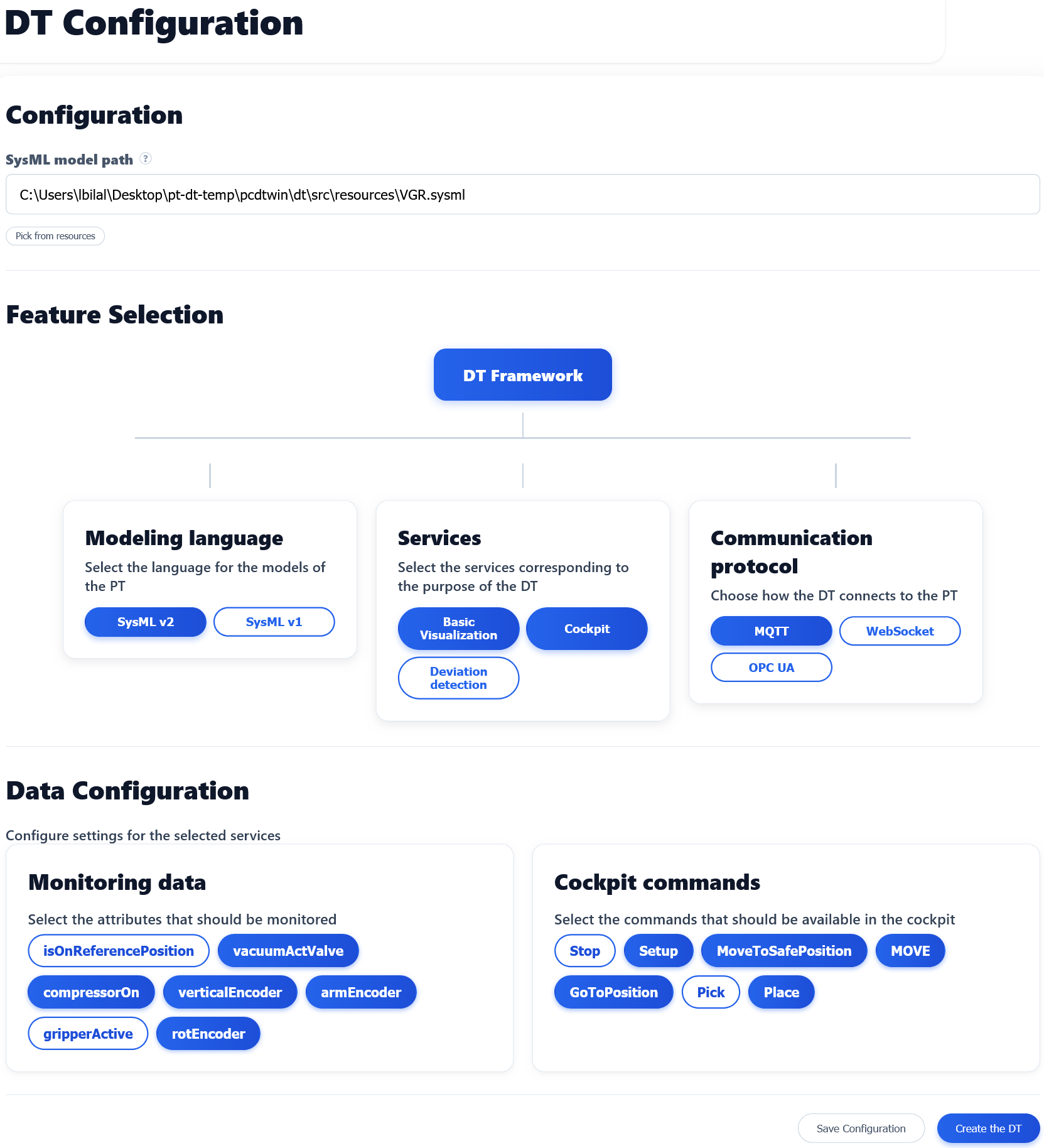}
\caption{Configuration User Interface for configuring the DT Framework.}
\label{configui}
\end{figure}

The framework is operated through a configuration UI (Figure \ref{configui}), that is intended for a domain expert. The configuration process starts by providing the design models of the PT, which serve as the basis for creating the DT. The domain expert then selects the configurable aspects of the DT, including the modeling language, the set of services that match the given purpose of the DT and the communication protocol to communicate with its physical counterpart. Depending on the selected services, the domain expert also selects which data should be visualized or which commands should be exposed through the cockpit. 

Once all the information has been entered and the domain expert clicks on \emph{Create the DT} button, the framework creates an instance of the DT. The DTEngine acts as the central orchestrator of this process and takes the configuration choices provided through the UI and uses them to set up the correct components. It uses the modeling language to activate the corresponding model interpreter, it instantiates the connectivity component for the chosen communication protocol so the DT can start receiving PT data and initializes the time series database used. During this initialization, the DTEngine applies the mapping between the model level information extracted from the SysML models and the incoming PT data, so that it can associate observed values with corresponding values defined in the models of the PT. It also activates the selected services and forwards to each service the configuration information it needs.

Communication between the DT and PT is managed through a gateway supporting multiple communication protocols. During the configuration phase, the domain expert selects the protocols to use, depending on the PT's available interfaces. Once the DT is running, the framework establishes the connection automatically with the selected communication protocols, subscribes to the PT data and forwards incoming events to the DTEngine.

Model interpretation is provided by the SysMLV2Interpreter, which parses the SysML models, and extracts both structural information and discrete behavioral specifications.
On the structural side, it exposes part definitions and their associated elements, including attributes, ports, actions and relevant constraints. 
This information is consumed by multiple components. It supports the configuration UI by exposing model defined data and actions available for selection and it supports the mapping layer by providing the structural model data that must be bound to incoming PT data. It also supports data validation by making constraints available when certain data must satisfy prescribed ranges or conditions. 

On the behavioral side, it evaluates incoming PT data by checking whether the values satisfy the guard conditions of defined transitions. When a match is found, the interpreter updates the DT's state accordingly. If any discrepancies are detected, whether it is a missing expected transition or an execution delay, the interpreter interacts with the anomaly detection service to flag the deviation.

As described in Section. \ref{sec:approach}, supporting an additional modeling language requires providing a new interpreter that can load and parse the model, extracting structural and behavioral elements that are exposed to the different components of the DT. 
Furthermore, the framework can support extensions for continuous behavior by interfacing with external simulation tools, allowing the management of both discrete and continuous behavior within the DT.

Real-time and historical data are stored in an InfluxDB\footnote{\url{https://www.influxdata.com/}} time-series database. Rather than relying on a predefined database schema, we initialize it from the structural elements exposed by the model interpreter. The framework creates InfluxDB measurements using the names of part definitions extracted from the structural models of the PT and associates the corresponding attributes to these measurements as fields, stored with their timestamps. The incoming PT data are written into this schema as provided by the mapping layer, so that each attribute defined in the models of the PT is continuously updated with its real time value.

\begin{figure*}[htpb]
    \centering
    \begin{subfigure}[t]{0.49\textwidth}
        \centering
\includegraphics[width=\columnwidth]{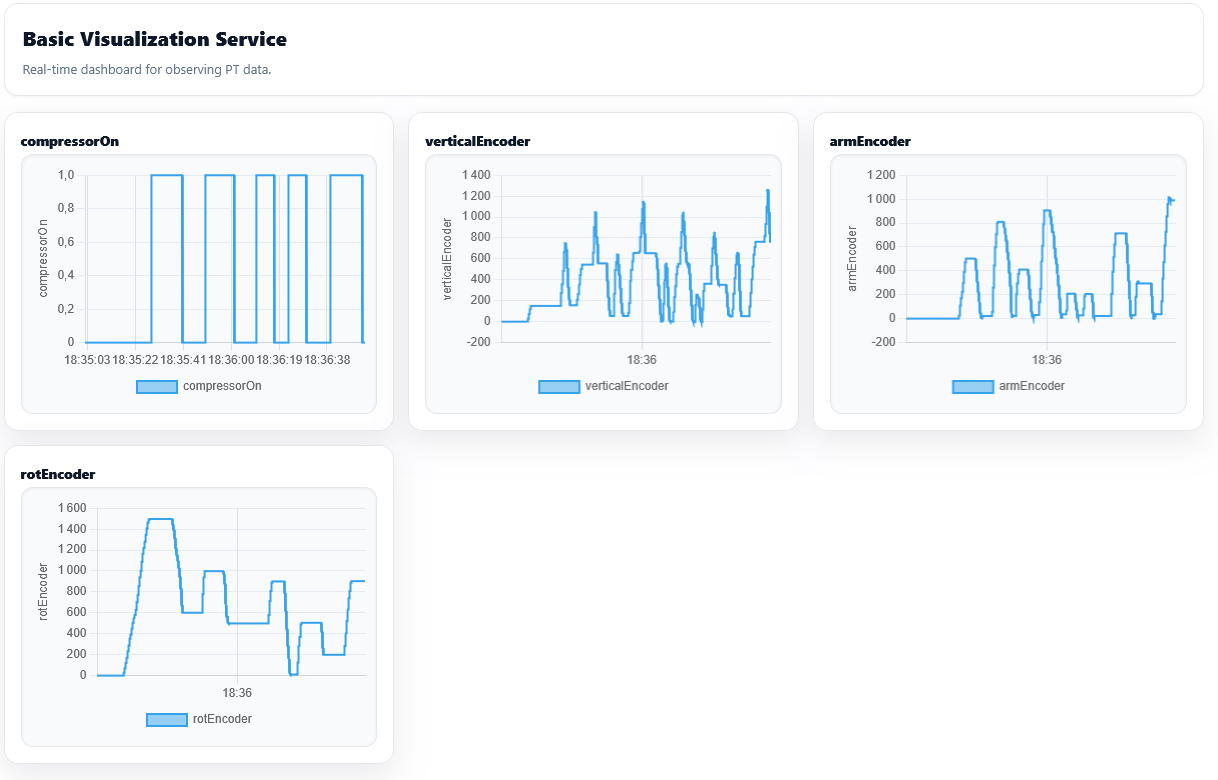}
\caption{Basic Visualization User interface with the selected data.}
\label{monitorUI}
    \end{subfigure}
    \hfill
    \begin{subfigure}[t]{0.49\textwidth}
        \centering
\includegraphics[width=\columnwidth]{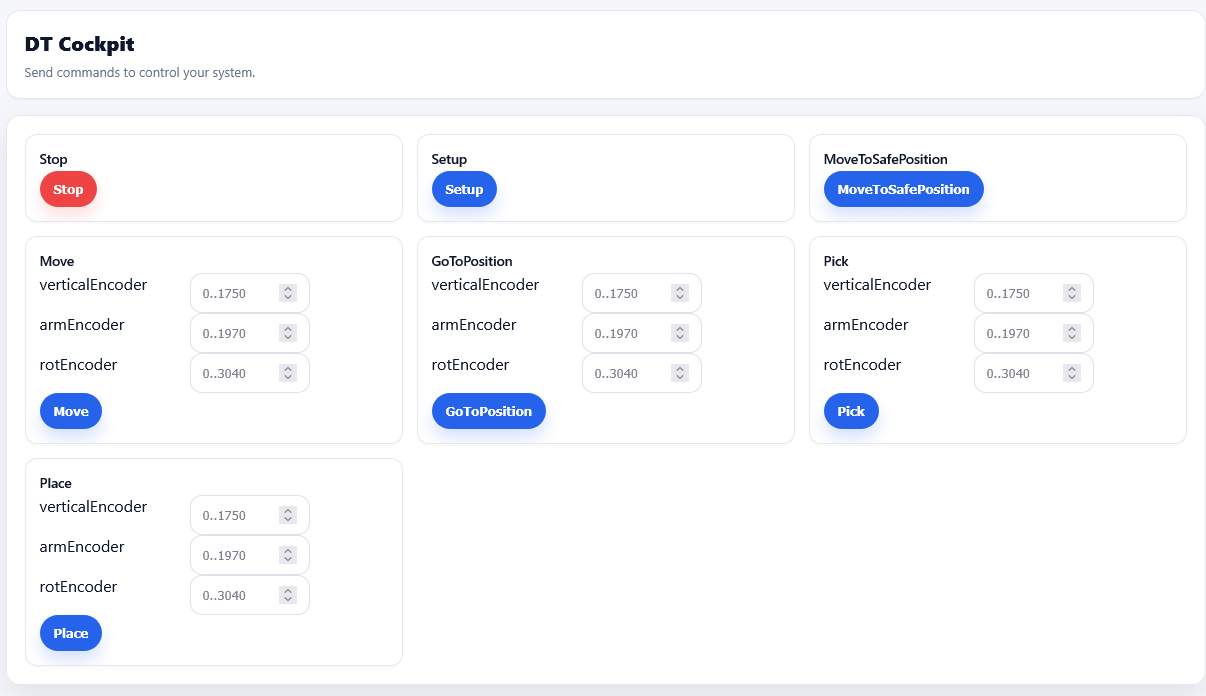}
\caption{Cockpit User Interface for sending commands to the PT.}
\label{cockpitui}
    \end{subfigure}
        \caption{Screenshots of the Basic Visualization Interface and the Cockpit User Interface.}
\end{figure*}

On top of this, the framework instantiates the services selected during configuration. The basic visualization service (Figure \ref{monitorUI}) provides a monitoring UI that visualizes the subset of attributes selected by the domain expert during configuration. 
The cockpit service provides a UI that exposes the selected actions as commands and allows sending them back to the PT through the gateway using the chosen protocol (Figure \ref{cockpitui}).

% \begin{figure}[htbp]
% \begin{center}
% \includegraphics[width=\columnwidth]{figures/monitoringUI.png}
% \caption{Basic Visualization User interface with the selected data.}
% \label{monitorUI}
% \end{center}
% \end{figure}

% \begin{figure}[htbp]
% \begin{center}
% \includegraphics[width=\columnwidth]{figures/cockpitUI.png}
% \caption{Cockpit User Interface for sending commands to the PT.}
% \label{cockpitui}
% \end{center}
% \end{figure}

Overall, this implementation demonstrates how each component of the framework is realized through concrete technologies and tools, and how their configuration and runtime behavior are driven by both the models of the PT and of the DT.

%% file: src/06.evaluation.tex
\section{Evaluation}

In order to evaluate our work, we examine to what extent our framework delivers on its intended benefits. We want to assess whether the automation of the DT development results in a reasonable configuration effort, such that a running DT can be obtained through configuration alone, based on existing models of the PT and with minimal additional development. We further evaluate the genericity of our framework by applying it to different use cases beyond industry 4.0 and analyse to what extent the approach can handle the co-evolution of the PT and DT.

\begin{figure*}[h]
    \centering
    \begin{subfigure}[t]{0.325\textwidth}
        \centering
\includegraphics[width=\columnwidth]{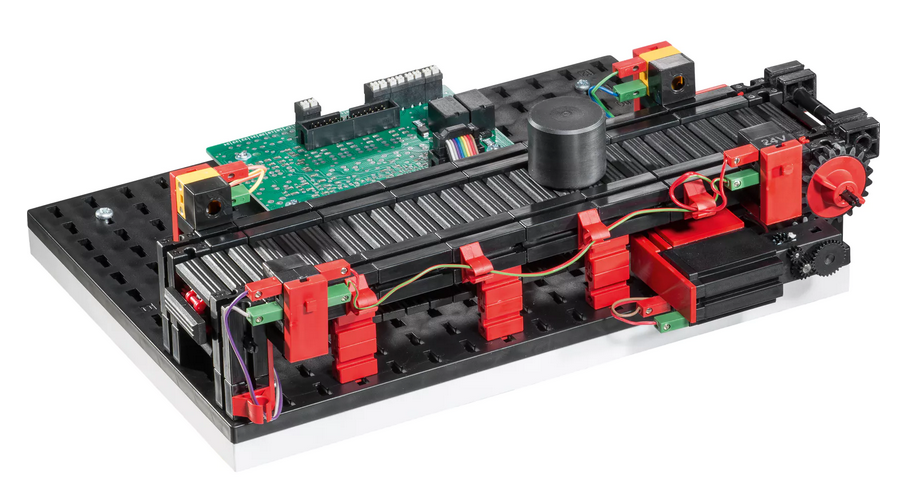}
\caption{UC1: Conveyor Belt of the Fischertechnik factory.}
\label{cbfig}
    \end{subfigure}
    \hfill
    \begin{subfigure}[t]{0.325\textwidth}
        \centering
\includegraphics[width=\columnwidth]{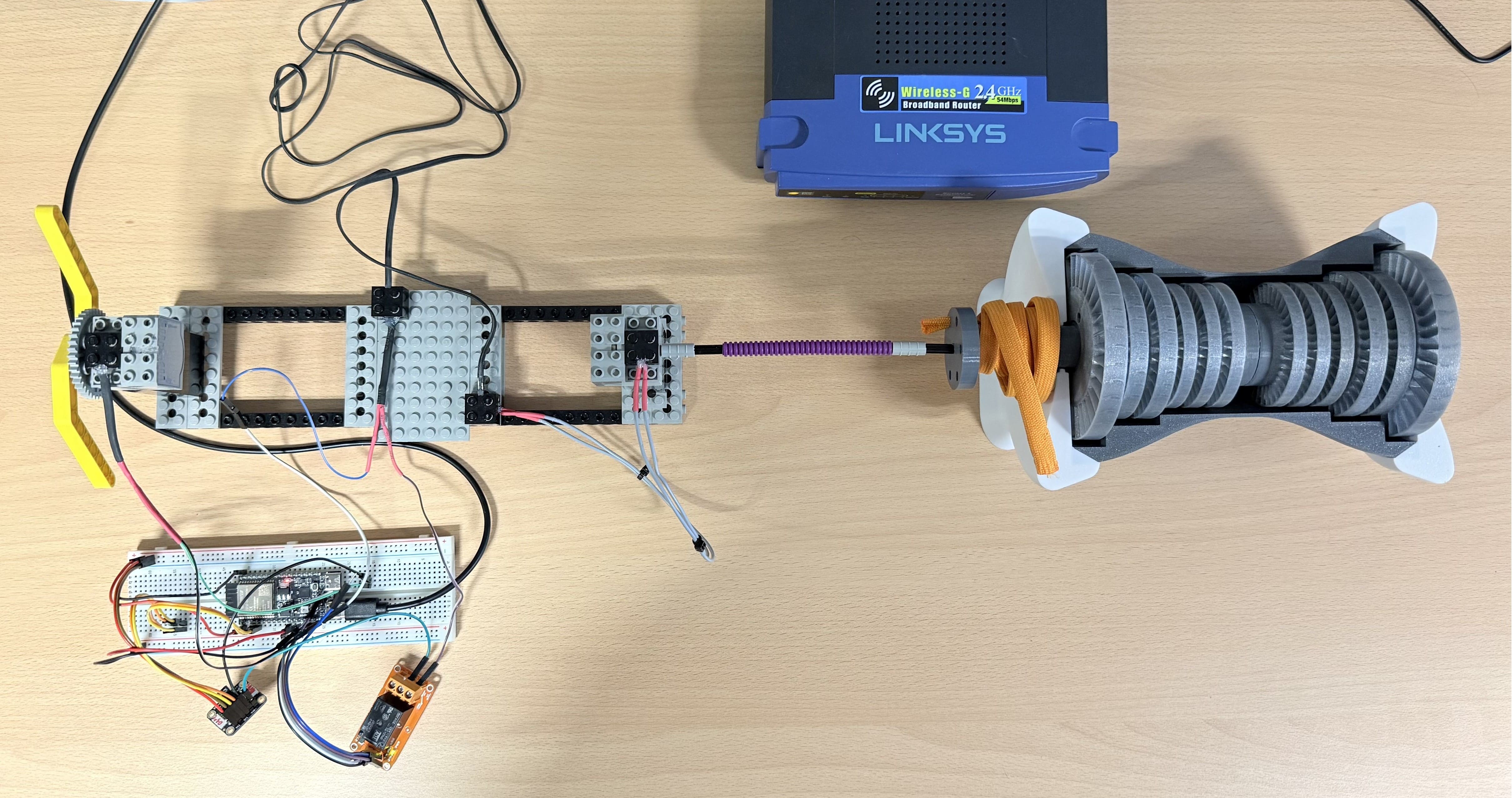}
\caption{UC2: Steam Turbine Demonstrator.}
\label{smfig}
    \end{subfigure}
    \hfill
    \begin{subfigure}[t]{0.325\textwidth}
        \centering
\includegraphics[width=\columnwidth]{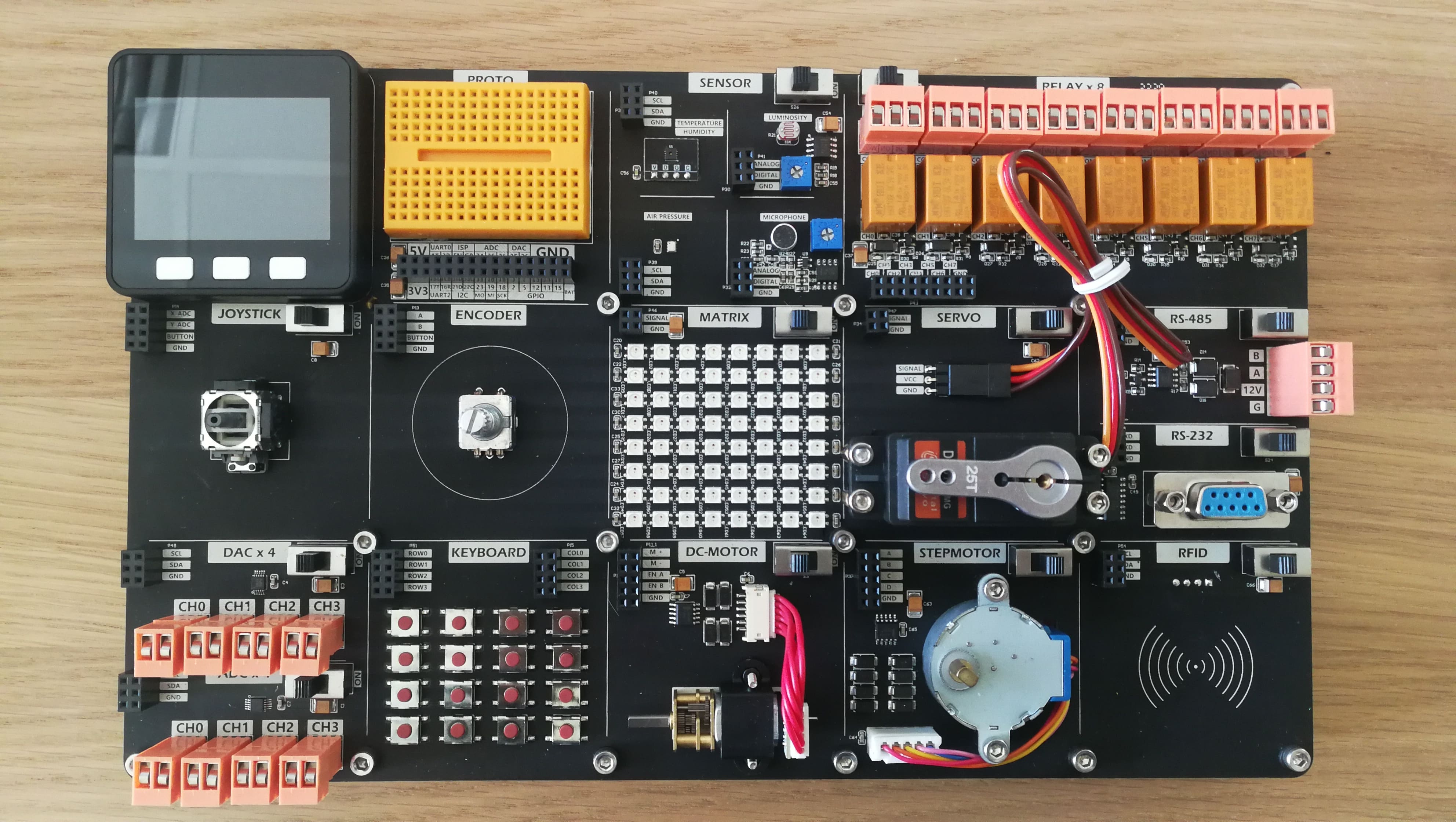}
\caption{UC3: Home Automation Demonstrator.}
\label{hafig}
    \end{subfigure}
    \caption{Evaluation use cases (UCs) excluding the VGR (UC0).}
    \label{fig:foobar}
\end{figure*}

\subsection{Research Questions}
We define the following research questions for the evaluation process of our framework.

\textbf{RQ1}: What is the implementation effort required at the Gateway level to connect the PT to the DT framework?

\textbf{RQ2}: To what extent can a domain expert obtain a working DT instance through configuration only? 

\textbf{RQ3}: To what extent can the framework be applicable across different use cases (UC)? 

\textbf{RQ4}: To what extent does the framework support the co evolution of the PT and DT over system changes? 

\subsection{\textbf{RQ1: Gateway Implementation Effort}}
To address RQ1, we quantify the effort required from the IIoT engineer by counting the lines of code in the gateway and protocol specific classes that integrate the PT with the DT framework and compare this to the LoC of our full DT implementation for the same system.

\textbf{UC0} corresponds to the VGR of the Fischertechnik demonstrator that was already described in the motivating example (cf. Section \ref{sec:ill}).  

For the VGR, the IIoT engineer only needed to implement the gateway logic that publishes PLC signals as MQTT messages and enriches these messages with high-level information about the commands, so that the DT can observe what the VGR is executing at each step. The PLC program that controls the VGR was already in place, so no changes were required on the PT side. This task required 168 LoC, compared to the 3903 LoC of our DT framework implementation, which corresponds to 4.3\% of the overall effort. 

\begin{tcolorbox}[colback=black!5!white,colframe=black!75!black]
The results show that the additional implementation effort at the gateway level remains small compared to our DT framework implementation.
\end{tcolorbox}

\subsection{\textbf{RQ2: Configurability Effort} }

To address RQ2, we conducted a configuration scenario on the VGR with ten participants, who had no prior experience in DT development but are familiar with modeling technologies, consisting of PhD students, along with one postdoctoral researcher and an engineer, to act as domain experts. The task was to obtain a working DT instance that (1) monitors a selection of sensors attributes via a dashboard in real-time, (2) exposes a cockpit to control the actuators and (3) detects anomalies in the behavior of the VGR. The participants interacted only with the SysML models of the VGR and the configuration UI, without writing or modifying the existent code. 
Before starting the configuration, we made sure these participants understood the use case and the demonstrator. We explained the purpose of the DT for this scenario and introduced the relevant SysML models so that they could understand them, as they are the basis for our DT framework. 

Once the task was clear, the participants began the configuration process by selecting the provided SysML models, choosing the configurable aspects of the framework, including SysMLV2 for the modeling language, the services needed to fulfill the DT goals as defined in the beginning of the experiment and the communication protocols already used by the demonstrator. 
They then proceeded to choose the data needed for the selected services, including which attributes to monitor and which commands to be exposed through the cockpit. For the anomaly detection service, the participants only had to select the corresponding service, since the DT reuses the SysML models to compare the expected behavior with the observed behavior of the VGR.

The task was completed within a few minutes, with an average configuration time of about 1 minute and 25 seconds, a maximum number of 20 actions and resulted in a running DT that displayed the selected attributes in the monitoring UI, exposed the selected commands in the cockpit UI and activated anomaly detection for the VGR.  

\begin{tcolorbox}[colback=black!5!white,colframe=black!75!black]
This end-to-end scenario captures the configuration effort required from a domain expert when using our framework, through models of the PT and configuration alone, with no changes to the framework code and the DT could be obtained in a matter of minutes.
\end{tcolorbox}

\subsection{\textbf{RQ3: Generalizability across UCs}}
To address RQ3, we apply our framework to 3 additional use cases. 

To demonstrate generalizability within the same factory, we instantiated the framework on an additional component of the Fischertechnik factory. \textbf{UC1} corresponds to the Conveyor Belt (CB) (Figure \ref{cbfig}) that is a transport system used to move workpieces along a belt. It is driven by a motor that can operate in both forward and reverse directions through PLC outputs. It is equipped with two photo transistors and light barriers to detect the presence of objects at the feed and swap stations, and a push button switch that is used to count the steps of the conveyor, incrementing each time the belt moves.
% \begin{figure}[htbp]
% \centerline{\includegraphics[width=\columnwidth]{figures/cb.png}}
% \caption{Conveyor Belt of the Fischertechnik factory.}
% \label{cbfig}
% \end{figure}

\textbf{UC2} corresponds to a steam turbine demonstrator (Figure \ref{smfig}) that consists of a steam machine and a generator, as used in a power plant. The PT is composed of a weight that simulates the action of steam for turbine rotation, which allows us to compute the potential energy at the turbine input. The turbine is coupled to a shaft connected to a generator that produces an electric current. The generator output is connected to an ESP32 microcontroller equipped with current and voltage sensors, which transmits data over Wi-Fi through MQTT. In addition, a relay controlled by the ESP32 via MQTT installed at the generator output simulates a connection point to the electrical grid at the power plant output. The electrical energy produced is then sent to a motor to simulate a load on the electricity grid. This use case focuses more on industry for energy production than only industry 4.0.

% \begin{figure}[htbp]
% \centerline{\includegraphics[width=\columnwidth]{figures/SM.jpeg}}
% \caption{Steam Turbine Demonstrator}
% \label{smfig}
% \end{figure}
Beyond Industry 4.0 applications, we also investigate the feasibility of using our approach on \textbf{UC3} that corresponds to a Home Assistant-based platform (Figure \ref{hafig}). The PT is a small home-automation demonstrator instrumented with an M5Stack Experiment Kit\footnote{\url{https://shop.m5stack.com/products/universal-iot-experiment-kit-for-esp32}} (ESP32). It is composed of a temperature and humidity sensor, a luminosity sensor, an air pressure sensor, a servo motor simulating a roller shutter, and an RGB LED matrix used as a controllable light source. The ESP32 publishes sensor readings via MQTT, and receives commands to control the actuators.
% \begin{figure}[htbp]
% \centerline{\includegraphics[width=0.7\columnwidth]{figures/home-assistant.jpg}}
% \caption{Home Assistant Domotic Demonstrator}
% \label{hafig}
% \end{figure}

For UC1 and UC3, we reverse engineered system models from the existing implementations so that they could serve as the basis for the corresponding DT instances. For UC2, we reused SysML models that had already been developed and used to build the PT.

% \begin{table*}[t]
% \caption{Comparison of Use Cases}
% \label{table-uc}
% \begin{tabularx}{0.98\textwidth}{
%   | >{\raggedright\arraybackslash}X
%   | >{\centering\arraybackslash}X
%   | >{\centering\arraybackslash}X
%   | >{\centering\arraybackslash}X
%   | >{\raggedright\arraybackslash}X
%   | >{\raggedright\arraybackslash}X |}
% \hline
% \textbf{Use case} & \textbf{Gateway Effort (LoC)} & \textbf{Effort (GW/DT)\% }  & \textbf{Services Used} & \textbf{Maximum number of actions}  & \textbf{Average number of actions} \\
% \hline
% \textbf{UC0 \,:\,VGR} & 168 & \(\approx 4.3\) & Monitoring \, Anomaly Detection \,  Cockpit & x & y\\
% \hline
% \textbf{UC1 \,:\,Conveyor Belt }&  146 & \(\approx 3.7 \) & Monitoring & x & y \\
% \hline
% \textbf{UC2 \,:\,Steam turbine} & 62 & \(\approx 1.6 \) &  Monitoring \, Cockpit & x & y \\
% \hline
% \textbf{UC3 \,:\,Home Assistant} & 119 & \(\approx 3.05 \) & Monitoring \, Cockpit  & 23 & y\\
% \hline
% \end{tabularx}
% \end{table*}

% Requires:
% \usepackage[table]{xcolor}
% \usepackage{multirow}

\begin{table*}[t]
\centering
\caption{Comparison of use cases.}
\label{table-uc}
\footnotesize
\begin{tabular}{>{\columncolor[HTML]{EFEFEF}}l|c|c|c|p{0.25\linewidth}|c|c}
\hline
\textbf{} &
\cellcolor[HTML]{A5D8FF}\textbf{\begin{tabular}[c]{@{}c@{}}DT Fram.\\ Implem. (LoC)\end{tabular}} &
\cellcolor[HTML]{FFF9DB}\textbf{\begin{tabular}[c]{@{}c@{}}Gateway Implem.\\ Effort (LoC)\end{tabular}} &
\cellcolor[HTML]{EFEFEF}\textbf{\begin{tabular}[c]{@{}c@{}}\% Effort\\ (GW/DT)\end{tabular}} &
\cellcolor[HTML]{EBFBEE}\textbf{\begin{tabular}[c]{@{}c@{}}Services used by the DT\end{tabular}} &
\cellcolor[HTML]{EFEFEF}\textbf{\begin{tabular}[c]{@{}c@{}}Maximum\\\#actions\end{tabular}} &
\cellcolor[HTML]{EFEFEF}\textbf{\begin{tabular}[c]{@{}c@{}}Average\\\#actions\end{tabular}} \\
\hline\hline
\textbf{UC0: VGR} &
\multirow{4}{*}{3903} &
168 &
$\approx 4.3$ &
Monitoring; Anomaly Detection; Cockpit&
20 &
8 \\
\textbf{UC1: Conveyor Belt} & &
146 &
$\approx 3.7$ &
Monitoring; Cockpit &
16 &
6 \\
\textbf{UC2: Steam Turbine} & &
90 &
$\approx 2.3$ &
Monitoring; Cockpit &
10 &
6 \\
\textbf{UC3: Home Assistant} & &
119 &
$\approx 3.05$ &
Monitoring; Cockpit &
23 &
9 \\
\hline
\end{tabular}
\end{table*}

Table \ref{table-uc} summarizes these use cases, the services used for each of them, the effort at the gateway level. The \emph{maximum number of actions} corresponds to the total number of interactions required to complete the configuration in the UI. The \emph{average number of actions} corresponds to the mean number of interactions required to obtain a running DT instance.

The gateway implementations vary across the different use cases, depending on the PT interface and on the amount of data to be exposed to the DT.

For UC1, the CB gateway follows the same structure as UC0 because both of the PTs are controlled through the PLC and expose comparable categories of signals. The implementation reuses the same broker configuration and publishing logic and differs mainly in the set of signals forwarded to the DT published. Since the CB exposes fewer signals than the VGR, its gateway implementation is smaller. 

For UC2, the steam turbine demonstrator requires a lightweight gateway that collects and publishes three kinds of measurements, related to the electrical production chain, including power, current, and voltage values and one command for turning on or off the demonstrator. The gateway also allows connecting or disconnecting the generator to the electrical grid using a relay. 

For UC3, the home assistant demonstrator relies on an ESP32 that publishes sensor readings and receives actuator commands over MQTT. The gateway therefore focuses on exposing a broader set of sensor attributes and actuators commands than in UC2. 

% (kind of answering RQ1 here also?) (paragraph was before moved use cases in this subsect. do we still leave it here? )Overall these results show that for our use cases, the gateway implementation remains small compared to the full implementation of the DT. The exact percentage can vary across use cases, depending on the abstractions handled by the gateway. In our scenarios, the gateway mainly forwards sensors values and actuator commands. If more complex processing were required, the effort would increase, but it would|could still remain significantly lower than re-implementing an entire DT from scratch. 

% --MENTION requirements here: - presence of sensors and actuators to read and write to them from DT - usage of an industrial communication protocol type mqtt, opcua and if other 
The complete set of artifacts for all use cases considered in this evaluation are available at \url{https://linaabilaal.github.io/MBDT-Framework/}.

\begin{tcolorbox}[colback=black!5!white,colframe=black!75!black]
These use cases make explicit the assumptions under which our framework can be applied. We assume that models of the PT are available, or can be reverse engineered, as these models are the basis for configuring the DT. 
The PT must also provide sensors and actuators that the DT can observe and control, via industrial communication protocols. 
% Additional protocols can also be supported, provided that appropriate protocol communication implementations are added to the framework and conform to its required interfaces. 
\end{tcolorbox}

\subsection{\textbf{RQ4: Co-Evolution of PT and DT}}

RQ4 investigates if our framework can support the co-evolution of the PT and its DT instance. We focus on evolution scenarios where sensors or actuators are added to or removed from the PT. In our approach, evolution is driven first at the model level, where the domain expert updates the SysML models of the PT, to reflect the changes, since these models are the basis for defining what the DT should observe and control. The IIoT engineer then applies the corresponding change on the actual PT and updates the gateway so that it starts or stops publishing the associated measurements. This update may require restarting the PT. During this time, the DT does not need to be stopped, it would remain running and wait for the PT connection and data streams to become available again.

% The domain expert adds a sensor or actuator to the PT, by modeling it into the models of the PT, the IIoT engineer then adds it on the actual PT and updates the gateway with the new information. The PT has to be relaunched, but the DT is designed such as they can handle the timeout until the connection is re-established.

On the DT side, these model updates are taken into account without restarting the DT. The updated attributes and actions appear in the configuration interface and become selectable. When the domain expert wants these additions to be reflected in a service specific view, they can trigger a reload action that refreshes the service UI.
% On the DT side, once the models of the PT are updated, without restarting the DT, the DT takes them into account, depending on the modification that happened at the model level.

\paragraph{Evolution Scenario 1: Adding a sensor}
\hfill\\
When a new sensor is introduced, the domain expert adds the corresponding attribute to the SysML model, the IIoT engineer would then add it onto the PT and update the gateway so that the new measurement is published. As soon as the gateway starts sending values, they are stored in the DS and become available to the DT services without interrupting its execution.
From the configuration perspective, the new attribute becomes visible and it can then be selected for the basic visualization service.

\paragraph{Evolution Scenario 2: Removing a sensor}
\hfill\\
When a sensor is removed, the domain expert deleted the corresponding attribute from the SysML model and the IIoT engineer removes it from the PT and stops publishing data from it. The attribute no longer appears in the configuration interface and the DT no longer considers it as part of the active configuration. Previously collected measurements remain stored for historical data for the DS.

\begin{tcolorbox}[colback=black!5!white,colframe=black!75!black]
These evolution scenarios show that the framework supports the co-evolution of the PT and DT with limited effort. Changes are handled through updates to the SysML models and at the gateway level, while the framework's code base remains unchanged, without needing to restart the DT. 
\end{tcolorbox}

% Since configuration is performed at design time, we consider that any modification to the models or to the configuration is taken into account at the next restart of the DT.
% - we tested two scenarios in our case, where we add a sensor, the DT takes into account and adds the newly attribute to the Digital Shadow and can see values coming from it. 
% if we remove a sensor, we would still like to keep it in the digital shadow for historical purposes [that aren't handled in our framework], but the DT takes it into account.
% The domain expert can select the added atttribute only at the next restart 

% Evolution Cases: remove a sensor equals removing an attribute from a certain block 

% add a sensor equals adding an attribute to the sysml models

% our DT framework is adapted to take into account the changes in the sysml models (so any modification whether added or deleted a sensor/actuator is taken into account) 
% since configuration is done design time, we decided for our framework that any modifcation on the configuration will be done at the next retstart of the DT, then the domain expert can take repick the added sensor|added actuator for example!

\subsection{Threats to Validity}

This section outlines potential threats to the validity of our evaluation and the steps taken to mitigate them.

\subsubsection{Construct Validity}

Construct validity relates to how well the evaluation metrics and methods reflect the concepts being studied. For RQ1, it is well known that using the number of lines of codes as a metric is a weak proxy for the effort needed for implementing the gateways. Still it gives a rough idea of the amount of work to be done.
From the IIoT engineer's perspective, the low gateway effort observed in our use cases is explained by the one-to-one mappings between PT data and model attributes. In scenarios where the gateway must implement more complex abstractions, such as data aggregation or complex event processing, the gateway implementation effort and the resulting effort may increase. Nevertheless, even in such cases, we expect the gateway effort to remain lower than the DT framework implementation.

\subsubsection{Internal Validity}
%  We ensured internal validity by having a full implementation of the experiment, tested several times, and recorded in a video that is publicly available \footnote{\url{https://youtu.be/pYUNYNtssDE}}. Still since our framework takes models of the PT as input, these models need to be accurate, as incomplete or imprecise behavioral definitions could lead to missed detections.
Internal validity concerns whether the observed outcomes are genuinely due to the system under evaluation rather than uncontrolled variables. Since our framework takes models of the PT as input, these models need to be accurate, as incomplete or imprecise behavioral definitions may lead to missed or incorrect deviation detections, independently of the framework's implementation. 

For RQ2, the configuration study involved a small number of participants, so the time of configuration is not really meaningful, which makes it difficult to draw definitive conclusions from the measured configuration times. Nevertheless, we still observed that a resulting DT instance could always be obtained in a matter of minutes without any prior knowledge of our framework.

\subsubsection{External Validity}
External validity addresses the generalizability of our findings. Beyond the VGR as a whole, we successfully applied the same method to the other use cases presented.
Our framework currently offers only three generic services, but in principle, additional services could be integrated into our framework if needed. The same applies to communication protocols and modeling languages, which can be extended.

We made the assumption that models of the PT were available, however even in our study, this was true for only one out of four use cases.evaluation. 
 In addition, for our approach to work, the models of the PT must be specified in SysML V1 or V2, as these are the formats supported by our framework.
To mitigate this limitation, models of the PT can be obtained by reverse engineering the PT they represent, although this adds additional effort that is not captured in our evaluation. evaluation. 
Also, if models of the PT are provided in other modeling languages, it is possible to extend our framework by adding a modeling language specific interpreter, otherwise, bridges can be built to SysML V1 or V2 through model transformations. 

We also do not evaluate scenarios where a large amount of data is removed from the digital shadow. Such changes can be costly and difficult to manage, as they may require specific data management policies to keep storage manageable. In our framework, we assume historical measurements remain available without further intervention.

In addition, the framework is built with support for protocol-specific connectivity and runtime interpreters for behavioral models, targeting systems that communicate via standard protocols and rely on deterministic state-based behaviors.
We currently focus on discrete behavior, which allows us to verify runtime alignment between PT and DT as required by the anomaly detection service. However, the framework does not support continuous behavior and therefore limits its predictive capabilities. Nevertheless, our framework could be extended to support continuous behavior when required.

However, our evaluation is mostly qualitative, and further quantitative analysis, as well as broader industrial scale validation are needed to strengthen the assessment of scalability and long-term maintainability, which will be addressed as future work.

%% file: src/07.relatedwork.tex
\section{Related Work}

In current literature, there are numerous articles that propose model-driven methods for developing DTs. 
%A systematic mapping study has been conducted by Lehner et al. \cite{lehner2025sms} to investigate how MDE automation techniques have been applied for DTs.

Bibow et al. \cite{bibow2020injection} present a model-driven approach to systematically develop DTs from models describing a cyber physical production systems (CPPS) and its domain, primarily focusing on defining reactive behaviors and specifying communication with the CPPS using domain specific languages. However, they rely on custom DSLs and a predefined reference architecture to derive the DT, whereas ours starts from already existing design models of the PT. Their solution also requires regeneration and redeployment of the DT when system models evolve, as it lacks support for runtime interpretation of dynamic updates.

Dalibor et al. \cite{Dalibor2020TowardsAM} propose a model-driven approach for systematically developing interactive DT cockpits, leveraging common data models and architecture modeling. However, their approach focuses more on the visualization and interaction with the PT from the cockpit. Their approach remains generic, focused mainly on visualization purposes, while we propose a generic framework adaptable for not only visualization, but also reasoning over models at runtime and controlling the PT. 

Kirchhof et al. \cite{kirchhof2020construction} propose a model-driven approach for the integration of cyber-physical systems (CPS) and DT information systems (DTIS). They automate software interface generation and maintenance, using Monti-Arc models for CPS architecture description and UML/P class diagrams for DTIS representation. While effective for system integration, their work predominantly targets the communication layer between the CPS and DTIS, while we target the whole architecture of the DT.

Lehner et al. \cite{lehner2023behavior} propose a pattern catalogue for integrating behavioral models into existing DT platforms, without modifying their code base. Their approach introduces modeling patterns that allow DTs to represent system behaviors and runtime execution alongside structural models. They demonstrate their method using Microsoft's DT platform on a 3D printer use case, highlighting how behavioral aspects can be integrated for simulation, validation and analysis. Nonetheless, their methodology emphasizes representing behavioral views within existing platforms rather than guiding the development process from design models. 

In \cite{michealwortmanndpdt,dalibor2022generationdt} a model-driven low code approach is proposed to support the creation and configuration of DTs, by allowing domain experts to choose the most suitable modeling languages for configuring and adapting DTs to their specific needs. Although their approach provides a flexible infrastructure for configuring DTs, it focuses primarily on DSLs tailored to the configuration and operation of DT services, while we reuse the models of the PT to configure the DT.

Munoz et al. \cite{munoz2021} explore the use of UML and OCL to model and test high level DTs. Their approach enables the specification and validation of DTs at a high level of abstraction, leveraging UML class diagrams and OCL constraints. However, they focus primarily on early testing and behavioral validation than on construction configuration of DTs.

Béchu et al. \cite{BechuBCPUV22} introduced a metamodel to provide concrete and operational descriptions for DTs deployment, specifically focusing on the definition of hardware and software components within layered cyber-physical architectures. This model-driven engineering approach defines a synthetic architecture for organizing software and data streams. The architecture organizes the DT elements into key packages, including the Core Twin, Data and History, Devices, External Tools, and Management components. %The proposed metamodel offers a generic, model-driven solution for constructing and deploying DT architectures, which was validated through two case studies: a Human-Robot Interaction system and a Home Automation System

Azangoo et al. \cite{azango2020dtmanu} also apply UML based modeling for DT engineering in a manufacturing context. They use class diagrams to model the DT and its interaction with PT. Their works emphasizes monitoring, fault detection and validation, through behavioral specifications, but relying on a unidirectional communication from the PT to the DT, without sending commands to the DT.
%and control capabilities to the DT. 

Lehner et al. \cite{aml4dtlehner} propose a framework for the model-driven development and maintenance of DT infrastructures, leveraging automation ML models. Their approach introduces a platform independent metamodel derived from existing DT platforms to maintain consistency across the different services.
While it addresses the automation aspect by generating DT models from engineering models, thus offering valuable insights on handling structural data in our framework, it does not address behavioral aspects, which are a key focus of our framework. Furthermore, they implement a digital shadow, since they do not focus on the the link from the PT to the DT. 

To the best of our knowledge, while there are existing works on MDE for DTs, there are currently no works on deriving a DT from models of the PT.

% In \cite{dalibor2022generationdt}, Dalibor et al. propose a two step approach for customizing low code develeopment platforms for creating and configuring DTs, combining auotmated generation of information systems with the modeling of DTs from architectural specifications, leveraging modeling platforms to define dt struccutres behaviors within a configurable platform. 

%% file: src/08.conclusion.tex
\section{Conclusion and Perspectives}
%persepctives: 
%- answer rq3 : 
%RQ3: What are the constraints between services on one side and models and data on the other side in a digital twin architecture ? 
 %( since we already choose in a way the data and services we want, we ca go further and actually see which models and data for which services and vice versa) 

In this article, we presented a tool-supported framework for automatically deriving a DT. To support automation and reusability, we reused models of the PT to derive certain aspects of the DT and we extended them with models of the DT to configure the framework. We demonstrated the feasibility of our framework on four different use cases.

As part of our future work, we intend to explore a multi-staging approach to determine which models and data are relevant for activating specific services. We also plan to explore the generalizability of our framework across different levels of abstraction in the models of the PT. This includes extending our feature model to account for abstraction levels and examining how they influence the resulting DT. Our models already reflect multiple abstraction levels, which positions our framework well for supporting configurable multi-fidelity DTs.

Another perspective is extending the framework with more specific services, rather than generic ones that were presented in the paper. In particular, a reconfiguration service could be supported by allowing domain experts define the desired reconfiguration rules through a high level language that is meaningful to them. This also supports the integration of specific services specified by the domain experts at design time.

%% file: src/09.acknowledgments.tex
\acknowledgments{
This work has been funded by the Agence Nationale
de la Recherche through the ANR-22-CE92-0068 Model-Based DevOps and by the Deutsche Forschungsgemeinschaft (DFG, German Research Foundation). Website: \url{https://mbdo.github.io}.
}